%Paper: hep-th/9301052
%From: Ruth Gregory <ruthie@yukawa.uchicago.edu>
%Date: Wed, 13 Jan 93 14:31:31 CST
%Date (revised): Fri, 15 Jan 93 11:58:56 CST

\magnification=\magstep1
\openup 2\jot
\def\bo{ { \sqcup\llap{ $\sqcap$} } }
\overfullrule=0pt       % delete the nasty little black boxes for overfull box

%\magnification=\magstep1

%\hsize=160truemm
%\hoffset=10truemm
%\rightskip 10truemm
%\leftskip 10truemm

\font\list=cmcsc10

\def\real{I\negthinspace R}

\def\r3{I\negthinspace R$^3$}

% HYPHENATION

\hyphenation{
mini-su-per-space
pre-factor
tei-tel-boim
es-po-si-to
haw-king
Bala-chand-ran
}

\hbox{ }
\rightline {EFI-93-02}
\rightline {January 1993}
\vskip 1truecm

\centerline{\bf BLACK STRINGS AND p-BRANES ARE UNSTABLE}
\vskip 1truecm

\centerline{Ruth Gregory}
\vskip 2mm

\centerline{ \it Enrico Fermi Institute, University of Chicago}
\centerline{ \it 5640 S.Ellis Ave, Chicago, IL 60637, U.S.A.}

\vskip 4mm

\centerline{ Raymond Laflamme }
\vskip 2mm

\centerline{ \it Theoretical Astrophysics, T-6, MSB288,
Los Alamos National Laboratory}
\centerline{ \it  Los Alamos,  NM 87545,USA}

\vskip 1.5cm
\centerline{\list abstract}
\vskip 3mm

{ \leftskip 10truemm \rightskip 10truemm

\openup -1\jot

We investigate the evolution of small perturbations around black strings
and branes which are low energy solutions of string theory. For simplicity
we focus attention on the zero charge case and
show that there are unstable modes for a range of time frequency
and wavelength in the extra $10-D$ dimensions.  These perturbations
can be stabililized if the extra dimensions are compactified to a
scale smaller than the
minimum wavelength for which instability occurs and thus will not affect large
astrophysical black holes in four  dimensions.  We comment on the implications
of this result for the Cosmic Censorship Hypothesis.

\openup 1\jot

}

\vfill\eject
\footline={\hss\tenrm\folio\hss}

Black holes are perhaps the most puzzling objects in general relativity.
They hide behind their horizon a singularity: a point which implies
the demise of the theory itself. The area surrounding this singularity is a
region of extremely strong gravity, and presumably is described by quantum
gravity.  In four spacetime dimensions black holes are stable,  once
formed they settle down to a state described solely by their mass, charge and
angular momentum, therefore the the singularities remain hidden from distant
observers.  This classical stability of black holes led to Penrose's Cosmic
Censorship Hypothesis$^1$ claiming that all singularities are hidden.
Quantum mechanically, black holes are quite different objects, they are
analogous to a thermal system.  The surface area of the hole
behaves like entropy, and
it is even possible to associate a temperature to a black hole as Hawking
has shown they radiate thermally$^2$.  However Hawking conjectured$^3$ that
a black hole formed from a pure quantum state would
radiate away leaving a mixed state
of radiation,   this would violate quantum mechanical unitarity.
It is difficult to understand the final stage of black hole evaporation as
general relativity is expected to break down at planckian curvatures, but
if quantum gravity preserves unitarity and information is to be
returned, it must do so well before the black hole reaches planckian
curvatures otherwise there is simply not enough energy left in a planck mass
black hole to emit all the information stored in a macroscopic black hole.

Recently, there has been a resurgence of interest in this problem,
largely due to the rise of string theory as a candidate for this unified
quantum theory. Many efforts have concentrated on the weak gravity r\'egime,
analysing the implications of low energy string theory on black hole structure.
Already some of these discoveries have been exciting. In Einstein gravity,
charged black holes (the Reissner-Nordstr\"om solutions) have an unfortunate
weakness. As well as an outer event horizon, they contain an inner Cauchy
horizon which is unstable to matter perturbations in the exterior spacetime.
However, there is no static charged black hole solution in Einstein gravity
with only one horizon and a spacelike singularity.
On the other hand, in low energy string theory,
gravity acquires a dilaton which greatly changes the causal structure of
charged black holes making them like Schwarzschild with one event horizon
and a spacelike singularity$^4$. This structure is generic, even if the
dilaton has a mass$^5$, as it must do to keep in line with the principle
of equivalence.
A particularly amusing aspect of these black holes is that in the extremal
limit of a magnetically charged black hole, the spacetime acquires an internal
scri at ``$r=2M$'' which is an infinite volume `throat' in which much
information can be stored.

Of course,
all of these models live in low dimensions, whereas string theory tells
us there are ten dimensions. Ideally therefore, one should be examining
black holes in ten dimensions. There has been work on black
holes in higher dimensions$^6$, including work that allows
for a range of horizon
topologies$^7$. In four dimensions, an event horizon must be topologically
spherical$^{8}$, but in higher dimensions this
is not necessarily the case, we
could have $S^2 \times $\real$^6$, or $S^3 \times $\real$^5$ topologies
for the horizon. The purpose of this letter is to point out that a large
class of these black holes are unstable under small perturbations. This is
a property which is very different from their analog in  four dimensions.
However there is a heuristic argument to show that this is reasonable.
Consider a five dimensional black string, Sch$\times$\real.
A portion of length $L$ has mass ${\cal M}=ML$
and entropy proportional to ${\cal M}^2/L$.
A five dimensional black hole on the other hand has
entropy proportional to ${\cal M}^{3/2}$. Thus for large lengths of horizon,
the mass contained within the horizon contributes a much lower entropy than
if it were in a hyperspherical black hole. This indicates that for large
wavelength perturbations in the fifth dimension, we might expect an
instability.

The issue of stability of the five dimensional black string
has been investigated analytically$^9$, with the result that
there is no non-singular single unstable mode
on a Schwarzchild time $t=0$ surface, however,
this argument did not prove stability.  As emphasized by Vishveshwara$^{10}$
in his
original Schwarzschild stability argument, the non-existence of a single
unstable mode does not preclude the existence of a composite unstable mode,
with the combination cancelling the singular behaviour of an inadmissable
single singular mode. This is in fact the situation with the coloured black
hole instability, recently confirmed by Wald and Bizon$^{11}$.
That this is indeed the situation for black strings was first indicated
by Whitt$^{12}$, who analyzed four
dimensional fourth order gravity and found an instability - a different
physical situation, but mathematically identical equations to
those studied in ref [9]. The key
simplification Whitt found useful was to use a different initial data
surface ending on the future horizon. By avoiding the neck of the Schwarzschild
wormhole, one avoids the fixed point of the isometries used to generate the
mode decomposition, which avoids in this case issues of superposition.
By adapting and generalizing his approach, we have been able to show that
extended uncharged black $p$-branes are unstable.
It is worth stressing that this instability is not of the Reissner-Nordstr\"om
form - hidden behind the event horizon, but it is a real {\it physical}
instability of the exterior spacetime which could potentially fragment the
horizon. It is important to emphasize that this can occur classically, for
although under regular conditions horizons do not bifurcate$^{13}$, if one has
a
naked singularity, then bifurcation is possible. Since an instability
calculation is by its nature linear, it cannot predict the endpoint of an
unstable evolution. However, the entropy argument
does lend support to the fragmentation scenario
and violation of the Cosmic Censorship Hypothesis.

In order to prove linear instability, an analysis of the perturbation equations
is required, with suitable reference to gauge and boundary conditions. Although
this process is quite detailed and involved, it is nonetheless possible to
present the salient features of the argument briefly. This is what we will now
do.

The black branes we are specifically interested in are those introduced
by Horowitz and Strominger in ten-dimensional low energy string theory
with a metric of the form
$$
ds^2 = - V dt^2 + {1\over V}dr^2 + r^2 d\Omega^2_{D-2} + dx^i dx_i .
\eqno (1)
$$
where $V= 1- (r_+/r)^{D-3}$, $D=4,..10$ and the index $i$ runs from
$1$ to $10-D$.
As we are only addressing uncharged black holes here, it is sufficient
to consider perturbations to the Einstein equations, since the dilaton
and gauge perturbations decouple and can be set to zero.
In the usual fashion we write a perturbation of the metric as
$$
g_{ab} \to g_{ab} + h_{ab}
\eqno (2)
$$
where we use the transverse trace-free (de Donder) gauge for $h_{ab}$:
$$
h^a_a = 0 = h^a_{b;a}
\eqno(3)
$$
This does not eliminate all of the gauge freedom, but does simplify the
perturbation equations
$$
\Delta_L h_{ab} = (\delta^c_a \delta^d_b \bo + 2 R_{a\;b}^{\;c\;d}
)h_{cd} \eqno (4)
$$
where $\Delta_L$ stands for the Lichnerowicz operator.

In general relativity, physics is invariant under general coordinate
transformations (gct's), which are generated by vector fields $\xi^a$.
The effect of an infinitesimal gct is to push the coordinates $\epsilon$
along the integral curves of $\xi^a$. Under such a gauge transformation,
the metric transforms as
$$
g_{ab}\to g_{ab} +2\xi_{(a;b)}
\eqno (5)
$$
hence a pure gauge perturbation of the metric is of the form
$$
h_{_\xi ab} = 2\xi_{(a;b)}
\eqno (6)
$$
But if $\xi^a$ is divergence free and harmonic, then $h_{_\xi}$ satisfies
both (3) and (4). Therefore, although there are $(N-2)(N+1)/2$
degrees of freedom in the solutions to the N-dimensional
Lichnerowicz equation, $(N-1)$ of
these are pure gauge, the remaining $N(N-3)/2$ being physical.
It will turn out to be fairly straightforward to identify the gauge degrees
of freedom.

Now, most importantly, there is the question of boundary conditions, which
are the key to this problem. Obviously, we want to place initial
data on a Cauchy surface for the exterior spacetime, but such a surface
necessarily touches the horizon, which is singular in Schwarzschild
coordinates. There are therefore two issues here: One is how to define
`small' for the perturbation at the horizon, and secondly, which initial
data surface to impose these constraints upon.

The first issue is straightforwardly dealt with. Although the horizon is
singular in Schwarzschild coordinates, it is not a physical singularity,
merely a coordinate singularity. In four dimensions, non-singular
coordinates have been known for some time - Kruskal coordinates. These
require generalizing to higher dimensions, which is slightly more involved,
but the transformation laws between Kruskal and Schwarzschild coordinates
remain qualitatively the same at the horizon. Therefore, since Kruskal
coordinates do not display their staticity in a straightforward manner,
we perform a mode decomposition in Schwarzschild coordinates, transforming
to Kruskal coordinates at the horizon to decide which modes are well
behaved.

This leaves us with the problem of an initial data surface. The domain
of dependence must obviously include ${\cal I}^+$, thus a surface touching
the future horizon, or the neck of the Schwarzschild wormhole is acceptable,
but a surface touching the past horizon is not, unless it passes through and
extends to the opposite horizon on the Penrose diagram. We choose the data
surface ending on the future horizon, as depicted on Figure 1,
for two reasons. One is that it
avoids the issue of mode superposition discussed earlier, and secondly, we
believe it to be a better physically motivated choice of surface. This is
because in practice a black hole (or brane) would form in a collapse
situation, and hence would not have a Schwarzschild wormhole; analyzing
stability would necessarily require a surface ending on a future event
horizon.

Now we turn to the actual stability analysis: are there any unstable modes?
Due to the symmetries fo the spacetime, we can split up the perturbation
into a purely transverse piece, a mixed transverse/D-Schwarzschild piece,
and a purely Schwarzschild piece.
This can be represented schematically as
$$
\left [ {\matrix{ h_{\mu\nu} & h_{\mu i} \cr
		   h_{j\nu}  & h_{ij} \cr}} \right ]
\eqno (7)
$$
where $\mu$ runs from 1 to $D$.
In a Kaluza-Klein spirit, we can interpret these perturbations as
scalar, vector and tensor respectively with respect to the D-dimensional
Schwarzschild spacetime.

It is relatively straightforward to show that there are no unstable modes
with non-zero scalar or vector pieces meeting our criteria of being
well behaved at both infinity and the future event horizon. However, for a
D-dimensional $s$-wave of the form
$$
h^{\mu i} = 0 = h^{ij}
$$
$$
h^{\mu\nu} = e^{\Omega t} e^{i \mu_ix^i} \left [
{\matrix{ H^{tt} & H^{tr} & 0 & 0 & .......\cr
	  H^{tr} & H^{rr} & 0 & 0 & .......\cr
	  0      &   0    & K & 0 & .......\cr
	  0  & 0 &  0 & {K\over \sin^2\theta} & ........\cr
	  ..      & ..     & .. & .. & .....\cr}} \right ]
\eqno (8)
$$
for certain values of $\Omega$ and $\Sigma \mu^2_i$, a solution to the
Lichnerowicz equation exists.

Using the metric (1) and the perturbation in the form (8) the Lichnerowicz
equation reduces to
$$
(\Delta^D_L + \sum_i\mu^2_i )h_{\mu\nu}  =0
\eqno (9)
$$
where $\Delta^D_L$ is the $D$-dimensional Lichnerowicz operator.  Note that a
pure D-dimensional gauge
perturbation, $h_{\mu\nu} = \xi_{(\mu;\nu)}$,
satisfies $\Delta^D_L \xi_{\mu;\nu}=0$.  Thus a
pure gauge perturbation of the metric must be a zero-mode
of the D-dimensional Lichnerowicz equation.
Therefore as long as $\mu^2 = \sum_i\mu^2_i \neq 0$ in equation (9),
$h_{\mu\nu}$ will be a real physical perturbation.

To find the equation obeyed by the perturbation we use the gauge
conditions to eliminate all but one variable from the
Lichnerowicz equation, $H^{tr}$ say, leaving a second order ordinary
differential equation:
$$
\eqalign{
0&= \Bigl \{
\textstyle{- \Omega^2 - \mu^2 V +
  {(D-3)^2 \left ({r_+\over r} \right)^{2(D-3)}\over 4r^2} }\Bigr \}{H^{tr}}''
-\Bigl \{
\textstyle{ {\mu^2[(D-2)- 2 \left ({r_+\over r} \right)^{D-3} +
  (4 - D) \left ({r_+\over r} \right)^{2(D-3)}] } }\cr
&\textstyle{ +{\Omega^2[(D-2)+(2D-7)\left ( {r_+\over r} \right )^{D-3} ]
   \over rV }
+ {3(D-3)^2\left ({r_+\over r} \right)^{2(D-3)}
  [(D-2)- \left ({r_+\over r} \right)^{D-3} ] \over
   4r^3V } }
\Bigr \} {H^{tr}}' \cr
&+ \Bigl \{ \textstyle{ \left( \mu^2 + \Omega^2 /V \right) ^2
+{\Omega^2 [ 4(D-2) - 8(D-2) \left ({r_+\over r} \right)^{D-3} -
       ( 53 - 34D + 5D^2 ) \left ({r_+\over r} \right)^{2(D-3)} ]
\over 4r^2V ^2} } \cr
& \textstyle{ +{\mu^2[ 4(D-2) - 4(3D-7)\left ({r_+\over r} \right)^{D-3} +
 ( D^2 +2D -11 ) \left ({r_+\over r} \right)^{2(D-3)} ]
 \over 4r^2V }}\cr
&\textstyle{ +{(D-3)^2 \left ({r_+\over r} \right)^{2(D-3)}
[(D-2)(2D-5) - (D-1)(D-2) \left ({r_+\over r} \right)^{D-3}
 + \left ({r_+\over r} \right)^{2(D-3)}] \over
4r^4 V^2}
} \Bigr\}  H^{tr}\cr
}
\eqno (10)
$$
By inspection of this equation,
the regular solution at infinity is $e^{-\sqrt{
\Omega^2 + \mu_i^2} r}$, and the solutions at the horizon behave as
$(r-r_+)^{-1\pm r_+\Omega/(D-3)}$.
Our boundary conditions demand the positive
root and $\Omega > 0$.
For $\Omega > (D-3)/r_+$ and any value of $\mu$, we can rule
out the existence of instabilities analytically.
%The behavior at infinity implies that the second derivative of $H^{tr}$ is
%postive.  In order to have a regular solution,  $H^{tr}$ must go
%through at least one local extrema.   At this point we must have a vanishing
%first derivative and a negative second one. This is impossible as neither of
%coefficients of the second and zero derivative change sign for
%$\Omega > (D-3)/r_+$.
Unfortunately it is exactly when $\Omega$ is of the order of $1/r_+$ that we
expect a possible instability.
For small $\Omega$ and
$\mu_i$ we can confirm the existence of regular unstable solutions numerically.
Obviously because the horizon is singular this process is delicate, however, if
we integrate in a regular solution from infinity, the general solution near
the horizon will be
$$
A_+(\mu) (r-r_+)^{ -1 + r_+\Omega/(D-3)}
+ A_-(\mu) (r-r_+)^{-1 - r_+\Omega/(D-3)}
\eqno (11)
$$
By taking appropriate combinations of this function and its derivative, we
determine the ratio $R = A_- / A_+$. Existence of a solution (and hence an
instability) is determined by a zero of $R$. We observe this in practice by
a change in sign of $R$, for which $R$ decreases as we home in on the sign
change. An increase in $R$ would indicate a zero of $A_+$. We found that
there did indeed exist zeros of $R$ for a range of $\Omega$, for all
$4\leq D\leq 9$, with appropriate values of $\mu$
shown in Figure 2.   The points in Figure 2 correspond to the values
calculated numerically and the lines have been added to guide the eye.

We have used the adaptive stepsize control of the Runge-Kutta routine described
in ref [14] to integrate the equation for $H^{tr}$  from a value of
$r1=200.0$ towards  $r_+$ stopping at $r2=2.0000002$. At this value we
calculated the ratio $R$.  The integration tolerance was set to
$eps=10^{-10}$.  We varied the integration parameters to check
the stability of the numerics.  There is a symmetry in the equation for
$H^{tr}$ under the following transformation: $r_+ \rightarrow \alpha r_+$,
$\Omega \rightarrow \Omega/\alpha$ and $\mu\rightarrow \mu/\alpha$, for a
constant value of $\alpha$.
Thus it is sufficient to calculate $\Omega$ and $\mu$ for only one value
of $r_+$.

The significance of these results is easily summarized:
black strings and branes are
classically unstable. This a real instability, for clearly the perturbation
cannot be written as pure gauge. By exhibiting a {\it single} $(\Omega,\mu)$
for any black brane, we prove instability, by exhibiting a range, we indicate
the instability is generic and robust.
How might we interpret this result physically? Of course, since our calculation
is linear, we cannot strictly say anything about the final state, but the
entropy argument, as well as the fact that $h_{ab}$ dominates $g_{ab}$ in
Schwarzschild coordinates near the horizon, makes it tempting to suggest that
the black brane will fragment. Periodic black hole solutions are known$^{15}$,
so there is a known final state solution in this case (unlike
Reissner-Nordstr\"om). Such a process will produce a naked
singularity and hence violate cosmic censorship. Perhaps a more realistic
though less spectacular conclusion  is that due to this instability, black
strings and p-branes will not form in the first place from collapse.

The only way around the instability is to compactify the transverse dimensions
on a scale smaller than the inverse mass of the black hole.
The compactification  would imply
that the values of $\mu_i$ are quantized. If their first value is greater
than the maximum one in Figure 2 this would imply that such `black
doughnuts' would be stable. Since there must be compactification of any extra
dimensions on an extremely small scale, all but the tiniest black doughnuts
would be safe, and those that would not would presumably have evaporated
producing their own naked singularities long ago.  Thus this instability
will  have no effect for contemporary astrophysical black holes.

Naturally this work makes no statement about classically charged black holes.
An investigation into these is in progress.
% and there is indication that they are also unstable
Nonetheless, the robustness of the
instability tempts us to conjecture that charged string and branes too
will be unstable for
some range of parameters, possibly tending to measure zero as the extremal
limit is approached. It does not seem to us that charge will prevent a black
brane from fragmenting, for the charge can collect on the individual black
holes maintaining overall charge conservation, though clearly destroying
any notion of charge per unit length.

If the charge is more topological in nature then some interesting things could
happen. For example, consider the axionic black holes of Bowick et.~al.$^{16}$
These
carry `quantum' charge, detectable only by an Aharanov-Bohm scattering process.
The field strength of the axion field is zero throughout the spacetime, but the
gauge field (rather like the gauge field of a local cosmic string) is
non-trivial due to the topology of the spacetime, $B = {Q\over \sin\theta
}d\theta \wedge d\phi$. This solution can clearly be extended to a five
dimensional black string (or ten-dimensional black 6-brane). This solution
too will be unstable, however, a five dimensional black hole cannot carry
the same type of axion charge. Drawing analogy to the gauge field of a local
cosmic string, it seems likely that during the fragmentation process, higher
energy physics could come into play, producing an axion vortex, which could
appear from behind the event horizon by an analogous process to the
t'Hooft-Polyakov monopole in four dimensions$^{17}$. The endpoint might then be
a line of black holes threaded by a cosmic axion string (not to be confused
with the four-dimensional global string).

Although the true endpoint of this instability is not presently known,
it could have important consequences for the Cosmic Censorship Hypothesis.
The form of $\delta g_{ab}$ indicates that these perturbations add
an oscillatory component to the location of the horizon as a function
of $x_i$ (the extra dimensions).
If these instabilities lead to a shrinking of the event horizon, black
holes singularities might reveal themselves.  A generic regular initial
perturbation  would therefore develop into a visible singularity.
The extremal case, where the event horizon and singularity coincide is of
particular interest.  If the event horizon shrinks, even by a very small
amount,
this instability may lead directly to naked singularity.  This case is
under present investigation.

Finally, to reiterate our original theme, this result makes clear the domain
of validity of four-dimensional Einstein gravity - namely, four-dimensional
Einstein gravity. The stability of four-dimensional Schwarzschild black
holes does not imply five-dimensional black strings or ten-dimensional black
branes are stable - indeed they are not! The result highlights the
unexpected subtleties of black holes, and is a demonstration that an event
horizon too can be ephemeral.

\proclaim  Acknowledgments.

We would like to thank J.Harvey, G.Horowitz, E.Martinec
and R.Wald for conversations.
R.G.  is supported by the McCormick fellowship fund at the Enrico Fermi
Institute,
R.L. appreciates support from Los Alamos National Laboratories.

\proclaim References.

\parskip=0pt

% MACROS

\newcount\refno
\refno=0
\def\nref#1\par{\advance\refno by1\item{[\the\refno]~}#1}

\def\book#1[[#2]]{{\it#1\/} (#2).}

\def\annph#1 #2 #3.{{\it Ann.\ Phys.\ (N.\thinspace Y.) \bf#1} #2 (#3).}
\def\cmp#1 #2 #3.{{\it Commun.\ Math.\ Phys.\ \bf#1} #2 (#3).}
\def\mpla#1 #2 #3.{{\it Mod.\ Phys.\ Lett.\ \rm A\bf#1} #2 (#3).}
\def\ncim#1 #2 #3.{{\it Nuovo Cim.\ \bf#1\/} #2 (#3).}
\def\npb#1 #2 #3.{{\it Nucl.\ Phys.\ \rm B\bf#1} #2 (#3).}
\def\plb#1 #2 #3.{{\it Phys.\ Lett.\ \bf#1\/}B #2 (#3).}
\def\prd#1 #2 #3.{{\it Phys.\ Rev.\ \rm D\bf#1} #2 (#3).}
\def\prl#1 #2 #3.{{\it Phys.\ Rev.\ Lett.\ \bf#1} #2 (#3).}

\nref
R. Penrose, \ncim 1 252 1969.

\nref
S.W.Hawking, \cmp 43 199 1975.
%black hole evaporation.

\nref
S.W.Hawking, \cmp 87 395 1982.
%loss of quantum coherence.

\nref
G.Gibbons and K.Maeda, \npb 298 741 1988.

D.Garfinkle, G.Horowitz and A.Strominger, \prd 43 3140 1991.

\nref
R.Gregory and J.Harvey, EFI 92-49, To appear in Phys.~Rev.~D.

J.Horne and G.Horowitz, UCSBTH-92-17, YCTP-92-P37.

\nref
R.Myers and M.J.Perry, \annph 172 304 1986.

\nref
G.T.Horowitz and  A.Strominger, \npb 360 197 1991.

\nref
W.Israel, \cmp 8 245 1968.

S.W.Hawking, \cmp 25 152 1972.

\nref
R.Gregory and R.Laflamme, \prd 37 305 1988.

\nref
C.V. Vishveshwara, \prd 1 2870 1970.

\nref
P.Bizon and R.Wald, \plb 267 173 1991.

\nref
B. Whitt, Ph.D Thesis, Cambridge, 1988.

\nref
S.W.Hawking and G.F.R.Ellis, \book The Large Scale Structure of Spacetime
[[Cambridge University Press 1973]]

\nref
W.H.Press, B.P.Flemming, S.A.Teukolsky and W.T. Vettering,
\book Numerical Receipes [[Cambridge University Press 1988]]

\nref
A.Bogojevic and L.Perivolaropoulos, \mpla 6 369 1991.

\nref
M.Bowick, S.Giddings, J.Harvey, G.Horowitz and A.Strominger, \prl 61 2823 1988.

\nref
K.Lee and E.Weinberg, \prl 68 1100 1992.

\bye